\documentclass[vecphys]{svmult}

\usepackage{makeidx}         % allows index generation
\usepackage{graphicx}        % standard LaTeX graphics tool
\usepackage{multicol}        % used for the two-column index
\usepackage[bottom]{footmisc}% places footnotes at page bottom
\makeindex             % used for the subject index
\usepackage{natbib}
\bibpunct{(}{)}{;}{a}{}{,}

\begin{document}

\title*{X-shooter: a medium-resolution, wide-band spectrograph for the
VLT}
\titlerunning{VLT/X-shooter} 

\author{L. Kaper\inst{1} \and
S. D'Odorico\inst{2} \and
F. Hammer\inst{3} \and
R. Pallavicini\inst{4} \and
P. Kjaergaard Rasmussen\inst{5} \and 
H. Dekker\inst{2} \and
P. Francois\inst{3} \and
P. Goldoni\inst{6} \and 
I. Guinouard\inst{3} \and \\
P.J. Groot\inst{7} \and
J. Hjorth\inst{5} \and
M. Horrobin\inst{1} \and
R. Navarro\inst{8} \and
F. Royer\inst{3} \and
P. Santin\inst{9} \and \\
J. Vernet\inst{2} \and
F. Zerbi\inst{10}}
\authorrunning{L. Kaper, et al.} 
\institute{Astronomical Institute, University of Amsterdam, The Netherlands
\texttt{lexk@science.uva.nl}
\and European Southern Observatory, Garching, Germany
\and GEPI, Observatoire de Paris, France
\and INAF, Osservatorio Astronomico di Palermo, Italy
\and Niels Bohr Institute, University of Copenhagen, Denmark
\and APC/UMR, Paris, France
\and Department of Astrophysics/IMAPP, Radboud University Nijmegen, The
Netherlands
\and ASTRON, Dwingeloo, The Netherlands
\and INAF, Osservatorio di Trieste, Italy
\and INAF, Osservatorio Astronomico di Brera, Italy}
%
% Use the package "url.sty" to avoid
% problems with special characters
% used in your e-mail or web address
%
\maketitle

X-shooter is the first second-generation instrument for the ESO {\it
  Very Large Telescope}, and will be installed in 2008. It is intended
  to become the most powerful optical \& near-infrared
  medium-resolution spectrograph in the world, with a unique spectral
  coverage from 300 to 2500~nm in one shot. The X-shooter consortium
  members are from Denmark, France, Italy, The Netherlands and ESO.

\section{X-shooter: a very efficient spectrograph}

The concept of X-shooter has been defined with one single main goal in
mind: The highest possible throughput for a point source at a
resolution which is just sky limited in about an hour of exposure
over the broadest possible wavelength range, without compromising
throughput at the atmospheric UV cutoff \citep{D'Odorico06}. The
moderate size of X-shooter is, as opposed to most existing or planned
VLT instruments, compatible with implementation at the Cassegrain
focus.

\begin{figure}[t]
\centering
%\hbox{
\includegraphics[width=12cm]{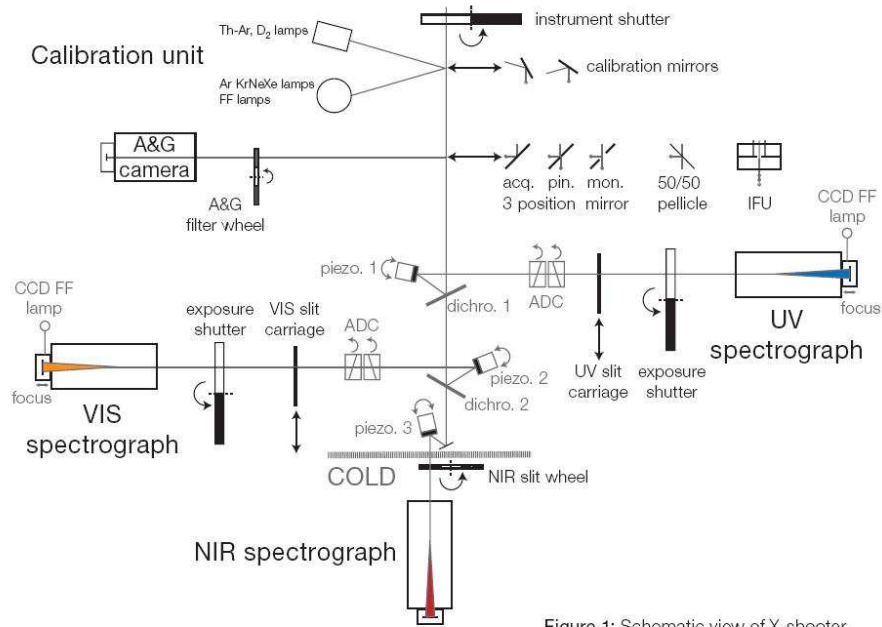}
\caption[]{
%{\it Left:} The X-shooter spectrograph depicted in the
%  Cassegrain focus below the M1 mirror cell of the VLT. To reduce
%  flexure, the center of gravity is located as close as possible to
%  the telescope, and the stiff elements are light-weighted. The UV and
%  VIS spectrograph arms are mounted to the side; the NIR arm is at the
%  bottom of the instrument; {\it Right:} 
Schematic view of X-shooter  \citep{Vernet07}.}
\label{fig1}       % Give a unique label
\end{figure}

The instrument design is based on multiple dichroics to split the
light between the three spectrograph arms (Fig.~\ref{fig1}). The
central backbone supports three prism cross-dispersed echelle
spectrographs (in double pass, optimized for the UV, visible and
near-infrared wavelength range and based on the so-called 4C design
\citet{Delabre89}). The backbone contains the calibration and
acquisition units, an IFU that can be inserted in the lightpath, the
two dichroics that split the light into the three arms and relay
optics to feed the entrance slits of the three spectrographs. The
spectral performance and efficiency (better than 95~\% reflectance,
resp. transmission) of the two dichroics (cross-over wavelength 550
and 1000~nm) turn out to be exceptionally good, especially when
considering the enormous wavelength range covered by X-shooter.

The standard slit measures $12'' \times 1''$; higher spectral
resolution is obtained when using the $12'' \times 0.6''$ slit
(Tab.~\ref{tab1}). A wide slit ($12'' \times 5''$) is available for
flux calibration. A dedicated program is being executed with
VLT/SINFONI to extend the calibration of 16 optical
spectro-photometric standards to the near-infrared in order to perform
flux calibration of X-shooter spectra aiming at an accuracy of better
than 5~\%.  The IFU has an entrance window of $4'' \times 1.8''$ and
delivers three slices filling a $12'' \times 0.6''$ exit slit.  The
Acquisition and Guiding unit has a $1.5' \times 1.5'$ field and includes a
comprehensive filter set; atmospheric dispersion compensation is
performed in the UV and VIS arms (though not when using the IFU).

\begin{figure}[t]
\centering
\includegraphics[width=12cm]{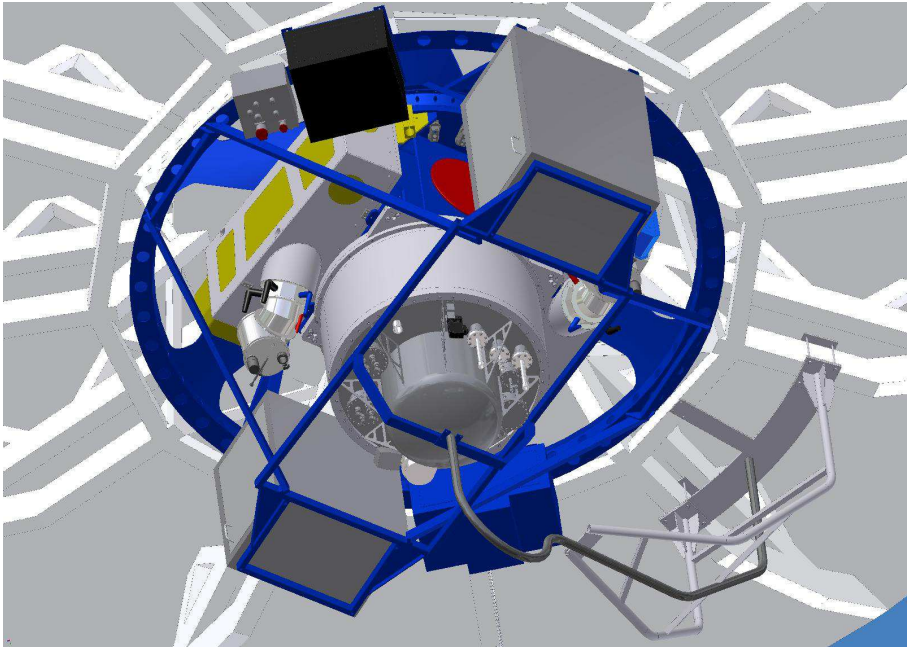}
\caption[]{The X-shooter spectrograph depicted in the
  Cassegrain focus below the M1 mirror cell of the VLT. To reduce
  flexure, the center of gravity is located as close as possible to
  the telescope, and the stiff elements are light-weighted. The UV and
  VIS spectrograph arms are mounted to the side; the NIR arm is at the
  bottom of the instrument.}
\label{fig2}       % Give a unique label
\end{figure}

Given its location in the Cassegrain focus, X-shooter has a tight
weight (less than 2.5 tonnes) and flexure budget
(Fig~\ref{fig2}). Measures have been taken to reduce the effects of
flexure, e.g. the optical bench of the NIR arm is extreme
light-weighted (machined from a block of 850~kg aluminium to a weight
of 25~kg, but similar strength). Also, three active flexure correcting
mirrors are added to the system. During target acquisition, a
calibration exposure is obtained from which the flexure is measured at
the current instrument position. This information is fed to the piezo
mirrors making sure that the object is centered on the three
spectrograph slits during the observation.

\begin{table}
%Tab. 1
\label{tab1}       % Give a unique label
\centering
\caption[]{Predicted average efficiencies (at the blaze of the echelle
  orders and outside the dichroic cross-over range) per spectrograph
  arm. The technical specification requires an efficiency $>
  25$~\%. Also the predicted spectral resolving power ($R = \lambda /
  \triangle \lambda$) is listed and compared to the technical
  specification $R_{\rm spec}$ for a $0.6''$ slit (excluding the
  detector LSF).}
\begin{tabular}{|l|c|c|c|c|}
\hline\noalign{\smallskip}
{\bf Spectrograph} & {\bf Spectral range} & {\bf Average blaze} &
{\bf $R_{\rm pred}$} & {\bf $R_{\rm spec}$} \\
               & (nm)   & {\bf efficiency} & ($0.6''$ slit) & ($0.6''$
slit) \\ \hline
UVB  & 307 -- 529 & 41.6~\% & 8169 & 7600 \\
VIS  & 558 -- 966 & 35.6~\% & 12335 & 11500 \\
NIR  & 1040 -- 2370 & 27.8~\% & 7329 & 7000 \\
\noalign{\smallskip}\hline
\end{tabular}
\end{table}

The near-infrared arm is cooled by liquid nitrogen. Originally a
closed-cycle cooler was planned to cool the instrument, but this could
induce vibrations on the telescope platform prohibiting VLTI
observations. The optical elements are cooled to 105~K, and the
optimal operating temperature of the near-infrared $2k \times 2k$
Rockwell Hawaii 2RG MBE detector is 81~K, just above the liquid
N$_{2}$ temperature (used area $1k \times 2k$).  The UV detector is a
$2k \times 4k$ E2V CCD; the VIS detector a $2k \times 4k$ MIT/LL
CCD.

Given the large wavelength coverage in one shot, a compromise has to
be found between the contribution of the read-out-noise in the UV
(requiring a long exposure time) and the sky background (variability)
in the near-infrared (short exposure time). This compromise likely
results in limiting the exposure time to about 20 minutes in ``staring
mode''. We are currently investigating whether nodding in the
near-infrared (using the telescope) can be compensated for in the
other two arms using the piezo mirrors.

About 75~\% of the costs of the X-shooter hardware, as well as labour,
is funded by the external members of the consortium. ESO is
responsible for the detector systems, project management, and
commissioning. More than 60 people are involved in the project at nine
different institutes distributed over four ESO member states and at
ESO. A complete list of the contributors to the project can be found
in \citet{Vernet07}. The overall cost of the project is 6.4~MEuro and
the staff effort 69~FTEs. The consortium is compensated for the
project investment with guaranteed time (156 nights over a
period of three years).  Even with the complex distribution of work
over many different sites, the X-shooter project has advanced well on
a relatively short timescale: $\sim 5$~yr from the official kickoff to
installation at the telescope.

\begin{figure}[t]
\centering
\includegraphics[width=12cm]{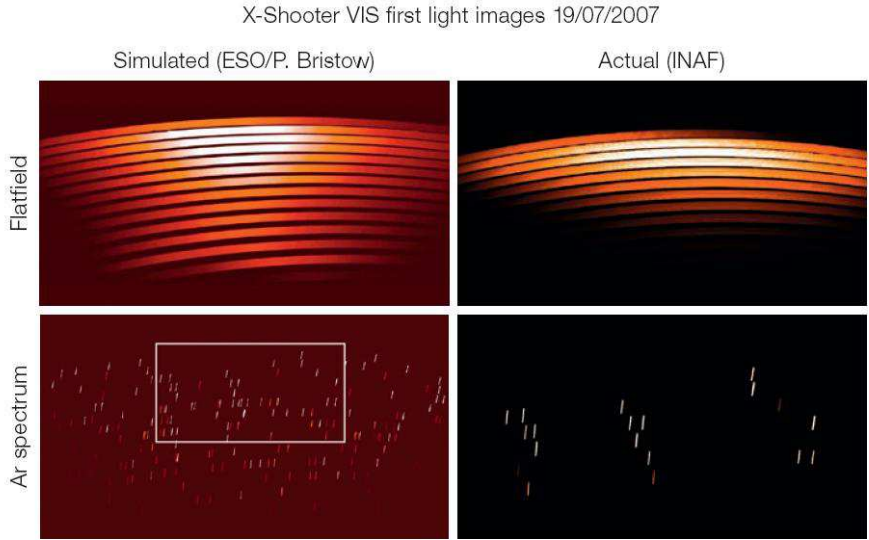}
\caption[]{Halogen flatfields and Ar line spectra in the VIS arm. The
  right panels show the first light images, the left panels display
  simulated images using the physical model. The first light Ar
  spectrum corresponds to the region enclosed by the white box in the
  simulated image.}
\label{fig3}       % Give a unique label
%\label{fig3}       % Give a unique label
\end{figure}

First light of the visual spectrograph was achieved in July 2007
%(Fig.~\ref{fig2})
(Fig.~\ref{fig3}); the NIR arm had first light in December 2007. In
January 2008 integration of the full instrument has started in ESO
Garching. The final delivery (the near-infrared arm) is planned for
March 2008. The system test phase will conclude with the so-called
Preliminary Acceptance Europe (PAE) review currently planned in June
2008. First light at the telescope is expected in September
2008. Normal operations should commence on April 1, 2009.

\section{Science with X-shooter}

X-shooter will have a broad and varied usage ranging from nearby
intrinsically faint stars to bright sources at the edge of the
Universe. The unique wavelength coverage and unprecedented efficiency
opens a new observing capacity in observational astronomy. At the
intermediate resolution of X-shooter 80-90~\% of all spectral elements
are unaffected by strong sky lines, so that one can obtain sky
continuum limited observations in between the sky lines within a
typical exposure time. Key science cases to be addressed with
X-shooter concern the study of brown dwarfs, the progenitors of
supernovae Type Ia, gamma-ray bursts, quasar absorption lines, and
lensed high-z galaxies. The advantage of the large wavelength coverage
is that e.g. the redshift of the target does not need to be known in
advance (such as in the case of GRBs); also, the study of
Lyman$\alpha$ in high-redshift galaxies will be possible in the
redshift range $1.5 < z < 15$.

X-shooter will complement and benefit from other major facilities in
observational astrophysics operational in the same period: survey
instruments like VST/OmegaCAM and VISTA working in the same wavelength
range, and observatories like ALMA, JWST and GLAST operational in other
observing windows.

\section{Expected instrument performance}

\begin{figure}[t]
\centering
\includegraphics[width=12cm]{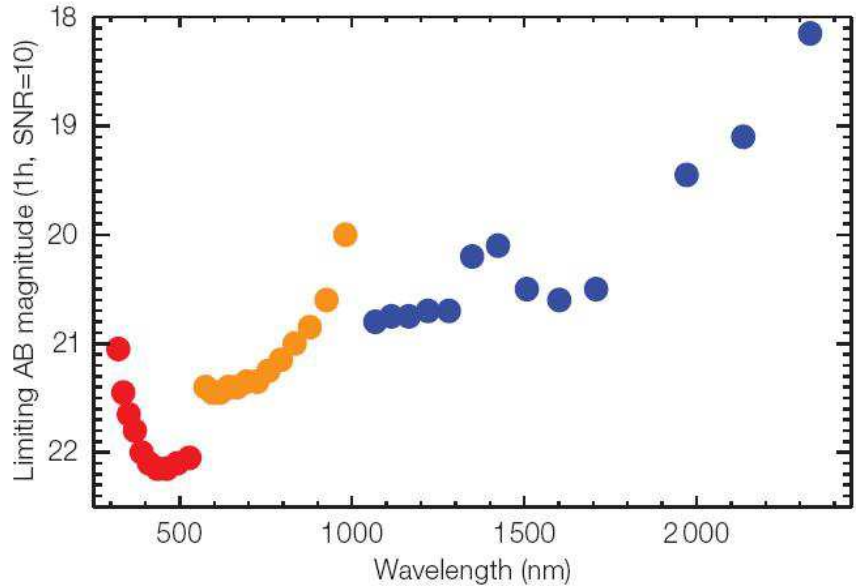}
\caption[]{Limiting AB magnitude of X-shooter per spectral bin at
  S/N=10 in a one-hour exposure. The other parameters used in this
  prediction are: airmass 1.2, 0.8$''$ seeing, 1$''$ slit, 2$\times$
  binning in the spectral direction. The first version of the ESO ETC
  was used to compute these values (www.eso.org/observing/etc).}
\label{fig4}       % Give a unique label
%\label{fig3}       % Give a unique label
\end{figure}

The performance of the instrument has been predicted on the basis of
measured efficiencies of the telescope, optical elements, and
detectors. Compared to the efficiencies predicted at the Final Design
Review in June 2006, most delivered optical components (most
importantly the dichroics and the gratings) are well above
specification: an efficiency $> 25$~\% in the centers of all orders
outside the dichroic cross-over range (from the top of the telescope
to the detector).  Table~1 lists the predicted efficiency at the blaze
of the echelle orders averaged over each of the three arms of the
instrument; the detection quantum efficiency is well over 40~\% in
most of the UVB range. Fig.~\ref{fig4}  
%Fig.~\ref{fig3} 
shows the predicted limiting AB
magnitudes (S/N = 10 per resolution element for a 1h exposure, from
the top of the atmosphere to the detector), calculated using a first
version of the ETC, assuming that the sky background is due to the
continuum in a region free of emission lines. In the near infrared the
exposure is split in 3 exposures of 20 minutes and nodding is
applied. The ETC uses the as-built values for optics and detector
efficiency/noise, but still contains some assumptions that need to be
verified during commissioning. The decrease in efficiency in the UV is
due to atmospheric absorption; towards the red there is a decrease in
CCD efficiency and the long-wavelength side of the near-infrared is
limited to the rise of the thermal background.  Also the predicted
spectral resolution complies with the technical specifications.

\section{Data reduction pipeline}

The X-shooter data reduction software is being developed as a
state-of-the-art ESO archival pipeline. About 15~\% of the
consortium's budget is spent on pipeline development, with the aim to
deliver spectra ready for scientific analysis. The pipeline contains
many novel features which are not commonly found in ESO pipelines,
such as: (i) full optimal extraction of data which is distorted in
both X and Y directions. The optimal extraction will be able to
automatically cope with arbitrary spatial profiles; (ii) The pipeline
will perform end-to-end error propagation. The pipeline includes a
physical model of the instrument (provided by ESO) that allows for
per-observation calibration of data, rather than relying on daytime
calibration data. This should deliver an absolute calibration accuracy
of 0.1 pixel for every frame taken by the instrument; (iii) A single
frame sky subtraction technique is being implemented, so that the
near-IR arm is useful for the long staring observations required by
faint targets at the shortest wavelength ranges of X-shooter.

%
%
% BibTeX users please use
% \bibliographystyle{}
% \bibliography{}
%
% Non-BibTeX users please follow the syntax
% the syntax of "referenc.tex" for your own citations
\bibliography{lkapervlt07.bib}
%%%%%%%%%%%%%%%%%%%%%%%%%%%%%%%%%%%%%%%%%%%%%%%%%%%%%%%%%%%%%%%%%%%%%%  }

%%%%%%%%%%%%%%%%%%%%%%%%%%%%%%%%%%%%%%%%%%%%%%%%%%%%%%%%%%%%%%%%%%%%%%

\printindex
\end{document}